\documentclass[twocolumn]{aastex61}

\usepackage{graphicx}
\usepackage{amsmath}
\usepackage{mathptmx}
\usepackage{times}
\usepackage{epsfig}
\usepackage{color}
\usepackage{mathrsfs}
\usepackage{amssymb}	% Extra maths symbols
\newcommand{\Mach}{\mathcal{M}}

\begin{document}

\title{On the use of hydrogen recombination energy during common envelope events}

%\title[Recombination energy and common envelope events]{On the use of hydrogen recombination energy during common envelope events}
\author[0000-0001-6251-5315]{Natalia Ivanova}
\affiliation{Department of Physics, University of Alberta, Edmonton, AB, T6G 2E7, Canada}

\correspondingauthor{Natalia Ivanova}
\email{nata.ivanova@ualberta.ca}

\begin{abstract}
  In this {\it Letter} we discuss what happens to hydrogen recombination energy that is released during
regular common envelope (CE) events, as opposed to self-regulated CE events.
We show that the amount of recombination energy that can be transferred away by either convection or radiation 
from the regions where recombination takes place is negligible.
Instead, recombination energy is destined to be used either to help CE expansion, as a work term,
or to accelerate local fluid elements. 
The exceptions are donors that initially have very high entropy material, $S/(k_B N_A)>37$ mol g$^{-1}$.
The analysis and conclusions are independent of specific stellar models or evolutionary codes,
and rely on fundamental properties of stellar matter
such as the equation of state, Saha equation and
opacities, as well as on stellar structure equations and the mixing length theory of convection.
\end{abstract}


\section{Introduction}

A common envelope event (CEE) is a fate-defining episode in the life
of a binary system.
The phenomena of a common envelope (CE) take place when the outer layers of one of the stars expand to engulf the companion \citep{Pacz76,1975PhDT.......165W,Webbink84,livio88}.
It is widely accepted that CEE is the dominant mechanism by which an initially wide binary star can either become a very close binary star, or by which it merges
\citep[for a review on the current understanding of the CE and its importance for binary populations,  see][]{Ivanova+2013Review}.
The outcome of a CEE depends on the energy budget during the interaction.

The foremost important topic in the consideration of the energy budget
during CE events is which energy can be, and which energy cannot be,
used to drive an envelope ejection. While the orbital energy release
is taken as the unarguable primary source, this becomes less efficient
as soon as the CE starts its expansion -- the expanded envelope
becomes tidally decoupled from a shrunken binary orbiting inside of it,
and the completion of the CE ejection using purely orbital energy is hindered
\citep{2016MNRAS.462..362I}.  Other energy sources that have been shown to
help to complete a CE ejection are nuclear energy released during the
interaction of the already shrunk binary \citep{2010MNRAS.406..840P}
and recombination energy \citep{2015MNRAS.450L..39N,2016MNRAS.460.3992N}.

The use of the recombination energy seems to be currently a subject
of controversy.  On the one hand, recombination energy
conveniently kicks in when the envelope is both expanded and decoupled
from the orbit, and if it takes place above the ``recombination radius''
\citep[for definition, see][]{2016MNRAS.462..362I},
it would provide enough energy to complete the ejection.
On the other hand, it has been argued
that the energy released by hydrogen recombination
is likely all transported to
the surface and then lost as radiation
\citep{2003MNRAS.343..456S,2017MNRAS.472.4361S,2018arXiv180305864G}.
The recombination energy is provided by both initially ionized hydrogen and helium, where helium provides about 60\% of the energy that hydrogen provides.
  For the recombination energy from helium, which is produced deeper inside an envelope,
  there is less controversy over its use.
  We hence will be not considering it in this {\it Letter}; 
  if hydrogen recombination energy cannot be transported, then
  helium recombination energy is not
  transported as well. 

For clarity,  we define the efficient use of  recombination energy
as when this energy is predominantly used locally to help drive a CE ejection, either by 
acting  as a  work term,  or being  converted into  kinetic energy  by
accelerating local fluid elements.  Note that we separate the issue of
the  efficiency of  recombination energy  usage in  regular CEEs  from that in 
self-regulated  CEEs,    also   known   as    slow   spiral-ins
\citep[first considered in][]{1979A&A....78..167M}. During  a self-regulated CE,  the envelope 
is allowed to readjust to match a moderate rate of orbital energy release
to its surface luminosity. We stress that this implies that
in  the case of self-regulated CEEs
the consideration of the efficiency of recombination energy is irrelevant
-- once the rate of orbital energy release is
matched by the surface luminosity, the recombination 
profile freezes.
\\
\section{Fundamentals}

The two fundamental stellar structure equations
that determine the energy redistribution inside a star are:

I. The energy transport equation\footnote{For equations 1-4 we refer the reader to appropriate texts, see, e.g., {\citet{2012sse..book.....K}.}}:

\begin{equation}
  \frac{L}{4\pi r^2} = F_{\rm rad} + F_{\rm conv}\ ,
  \label{eq:lum}
\end{equation}

 \noindent Here, $L$ is the luminosity at the mass coordinate $m$, $r$ is the radial coordinate, and $F_{\rm rad}$ and $F_{\rm conv}$ are the radiative and the convective flux, respectively, at the same mass coordinate.

 II. The energy equation. 
If there is no nuclear energy generation or neutrino losses, the net energy loss from a mass shell $dm$ is $dL$, and the energy equation is

\begin{equation}
  \frac{d L}{dm}=-\frac{\partial e}{\partial t}+\frac{P}{\rho^2}\frac{\partial \rho }{\partial t} 
  \label{eq:en_eq}
\end{equation}

\noindent Here  $t$ is time, $e$ is specific internal energy, $P$ is pressure, and $\rho$ is density.
The two terms on the right hand side of Equation~\ref{eq:en_eq}  are often combined into a gravitational energy term. We write it in two components to emphasize that during expansion, the second term acts as an ``energy sink'' due to work spent in expanding the shell, while the first term acts as a local energy source.

Equations~\ref{eq:lum} and \ref{eq:en_eq} show that the amount of energy that can be transferred away is limited by the fluxes. If the fluxes cannot transfer away the locally produced energy, then that energy has to be spent on the shell's expansion.  Hence, the intrinsic ability of the two fluxes to take the energy away determines the efficiency of recombination energy usage during a CE event.
In addition to convective and radiative energy transport mechanisms, some energy can be transported by waves \citep[see, e.g.][]{2017MNRAS.470.1642F}. Energy transported by waves  will be not lost from the surface, but rather deposited somewhere closer to the surface, and hence likely contributes to the  removal of the CE.

\

\noindent {\bf 2.1 The radiative flux}

The flux carried by radiation is

\begin{equation}
F_{\rm rad} = \frac{4 a c G}{3} \frac{T^4 m}{\kappa P r^2} \nabla = \frac{4 a c }{3} \frac{T^4 }{\kappa P } g \nabla 
\end{equation}

\noindent Here  $a$ is the radiation density constant, $c$ is the speed of light, and $G$ is the gravitational constant. $\nabla\equiv\frac{d \ln T}{d \ln P}$ is the actual gradient in the star; for radiative regions it cannot exceed $\nabla_{\rm ad}$. $T$ is temperature,  $\kappa$ is opacity, $g$ is gravitational acceleration. All those quantities are local to the considered shell at the mass coordinate $m$.  

\

\noindent {\bf 2.2 The convective flux}

The flux carried by convection within the mixing length theory (MLT) is 

\begin{equation}
F_{\rm con} = \rho v_{\rm conv} c_p  T (\nabla -\nabla_{e}) \frac{l_m}{2 H_p}
\end{equation}

\noindent Here $(\nabla -\nabla_{e})$ is the excess of $\nabla$ above the variation of temperature in the convective element during its motion, $\nabla_e$. This difference is limited by the maximum value of the temperature gradient in the case of adiabatic convection $\nabla_{\rm ad}$. This value is often 0.4,   but in a zone of partial ionisation it can become even less than 0.1.  $c_p$ is the specific heat at constant pressure. $l_m$ is the mixing length and $H_P$ is the pressure height scale, in MLT they are usually connected via the mixing length parameter $\alpha_{lm}=l_m/H_p$ which is often taken to be about 2. It is important that the MLT has been derived in the assumption that the convection is subsonic so that a convective eddy can always reestablish pressure equilibrium with its surroundings as it moves.  $v_{\rm conv}$, which is the velocity of a convective element, then can be considered as $v_{\rm conv}=\Mach_{cv} c_s$, where Mach number $\Mach_{cv} <1$ and $c_s$ is the local sonic velocity $c_s=\sqrt{\Gamma_1 P/\rho}$. $\Gamma_{1}=(\partial \ln P / \partial \ln \rho)_{\rm ad}$ is the first adiabatic exponent.   Currently, there is no stellar convection theory that would provide a valid result for a transonic or faster convection. We can now rewrite the convective flux as

\begin{equation}
  F_{\rm con} = \Mach_{cv }\rho c_{s} c_p  T (\nabla -\nabla_{e})
  \label{eq:fconv}
\end{equation}

\section{Ability of the radiative flux to remove the recombination energy}

\begin{figure}
        \includegraphics[width=3.6in]{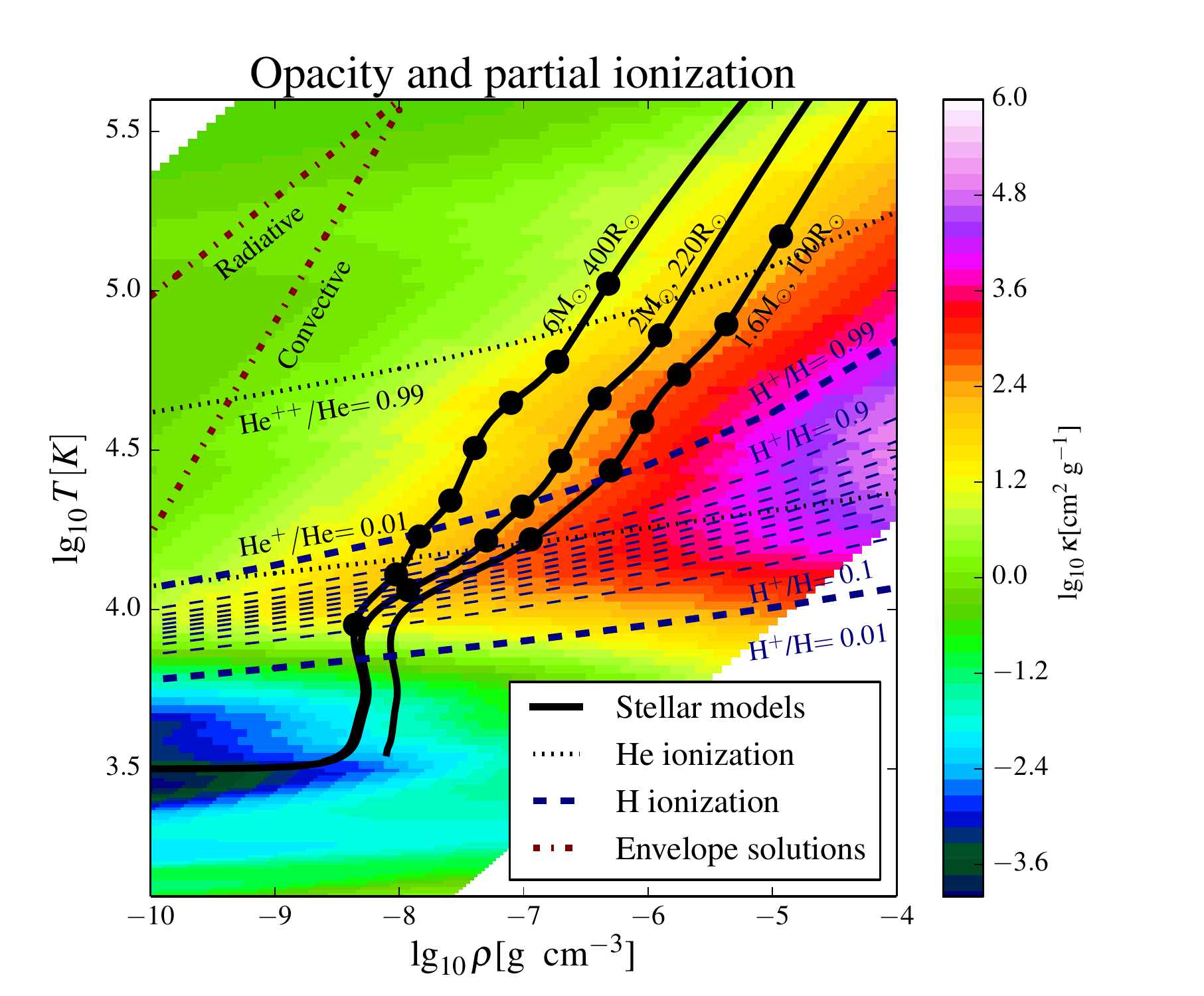}
        \caption{Envelope structures of three stellar models (AGB star with mass of 6 $M_\odot$ and radius of 400 $R_\odot$,
          AGB star with  mass of 2 $M_\odot$ and radius of 220 $R_\odot$, and RG star with mass of 1.6 $M_\odot$ and radius of 100 $R_\odot$) are plotted on a $T-\rho$
          (temperature-density) plot.
          The solid dots indicate where the mass above this location on the $T-\rho$ curve is $\delta M=0.01, 0.05, 0.1, 0.2, 0.4, 0.8, 1.6, 3.2 M_\odot$
          (all levels are applicable only for $6 M_\odot$ AGB star).
          Colors correspond to opacities from \citet{2005ApJ...623..585F} generated for metal abundances from \citet{1998SSRv...85..161G}, for $X= 0.7$ and $Z= 0.02$.
          Locations of Hydrogen partial ionization zones ($H^{+}/H=0.01,0.1,0.2,0.3,0.4,0.5,0.5,0.7,0.8,0.9,0.99$) and Helium partial ionization zones ($He^+/He=0.01$ and $He^{++}/He=0.99$) are found as in \citet{2012sse..book.....K}.
        The dash-dotted red lines (in upper left corner) indicate the inclinations that radiative and convective envelope solutions would have on the $T-\rho$ plane if radiative pressure is negligible, and there is no partial ionisation  ($\nabla_{\rm ad}=0.4$) -- above the recombination regions all the solutions are self-similar and are only shifted from each other. For details see \citet{2012sse..book.....K}. The shown stellar envelope profiles were obtained using {\tt MESA} code \citep{Paxton2011, Paxton2013, Paxton2015}, revision  10108.}
    \label{fig:op_ion}
\end{figure}
\begin{figure}
        \includegraphics[width=3.6in]{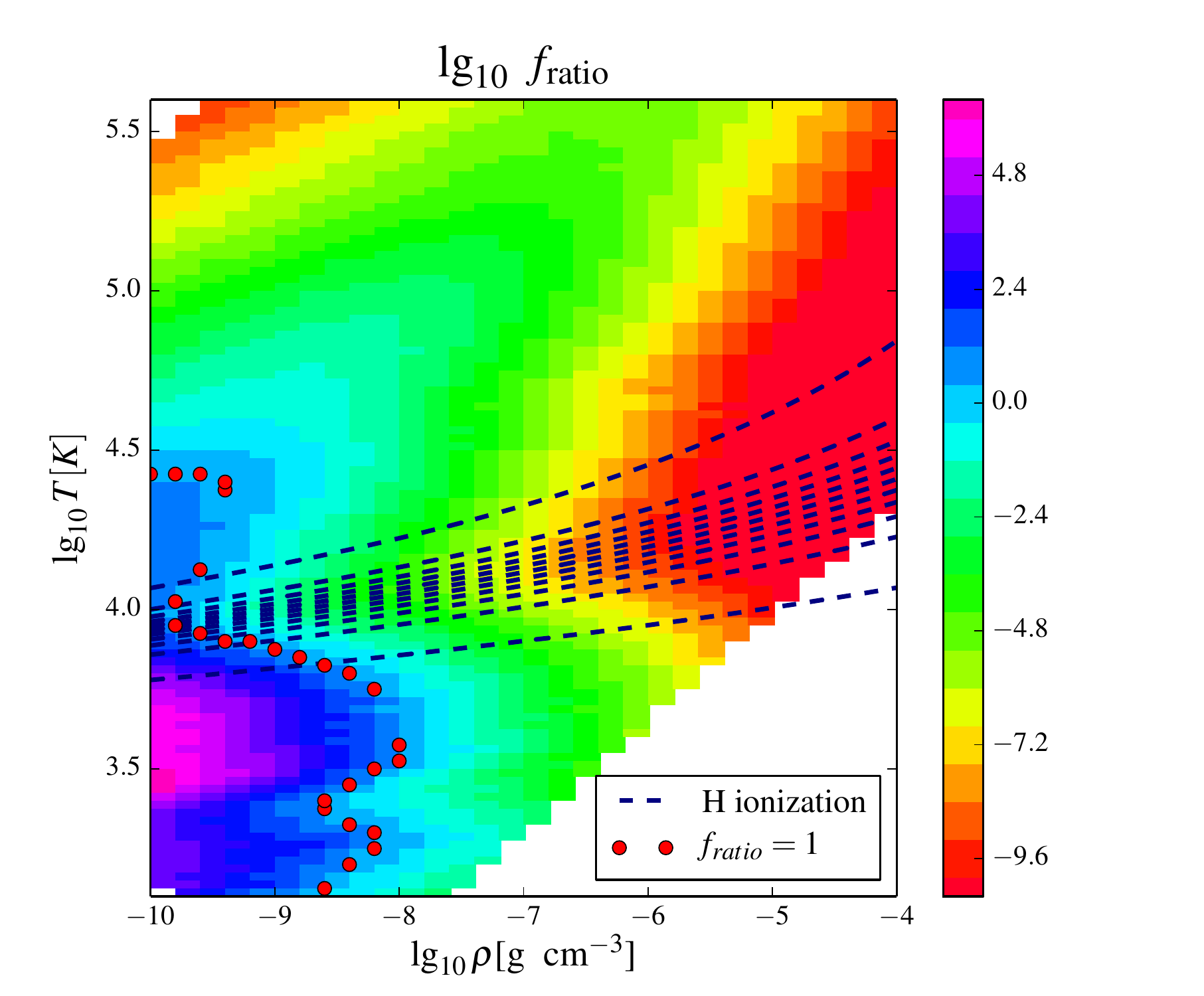}
        \caption{This figure shows $f_{\rm ratio}$ as defined by Equation \ref{eq:fratio} for a stellar mixture with $X=0.7$ and $Y=0.28$. Lines for $H$ partial ionization zones are as in Figure \ref{fig:op_ion}.}
    \label{fig:fratio}
\end{figure}
\begin{figure}
        \includegraphics[width=3.6in]{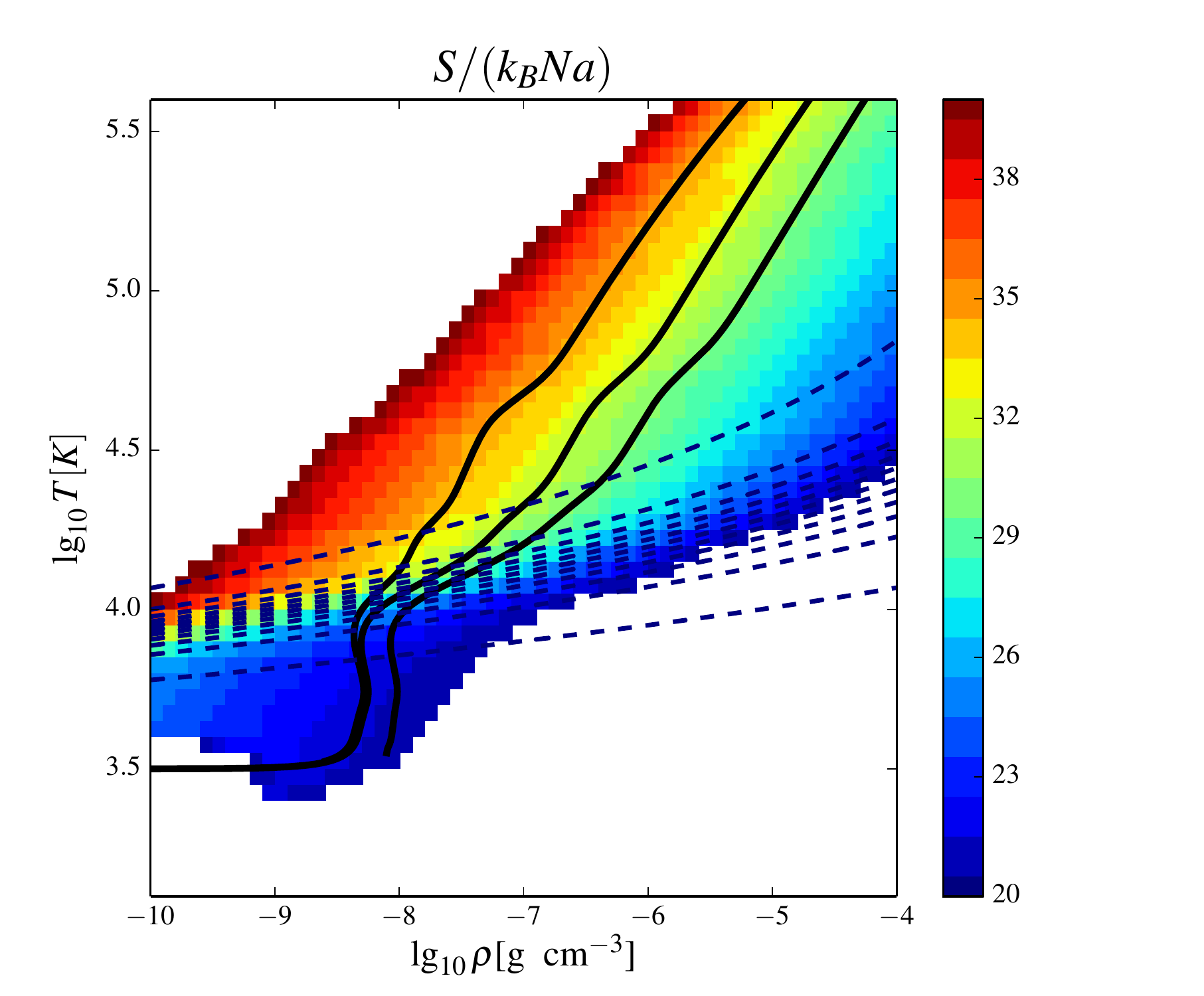}
        \caption{Scaled entropy $s=S/(k_B N_A)$ mol g$^{-1}$
          for a stellar mixture with $X=0.7$ and $Y=0.28$.
          Entropy values are cut below 20 and above 40 for better resolution in the region of interest;
          otherwise they are much higher at the left upper corner,
          and much lower in the right bottom corner of the plot.
          Lines for $H$ partial ionization zones and stellar models are as in Figure \ref{fig:op_ion}.}
    \label{fig:entr}
\end{figure}
\begin{figure}
        \includegraphics[width=3.6in]{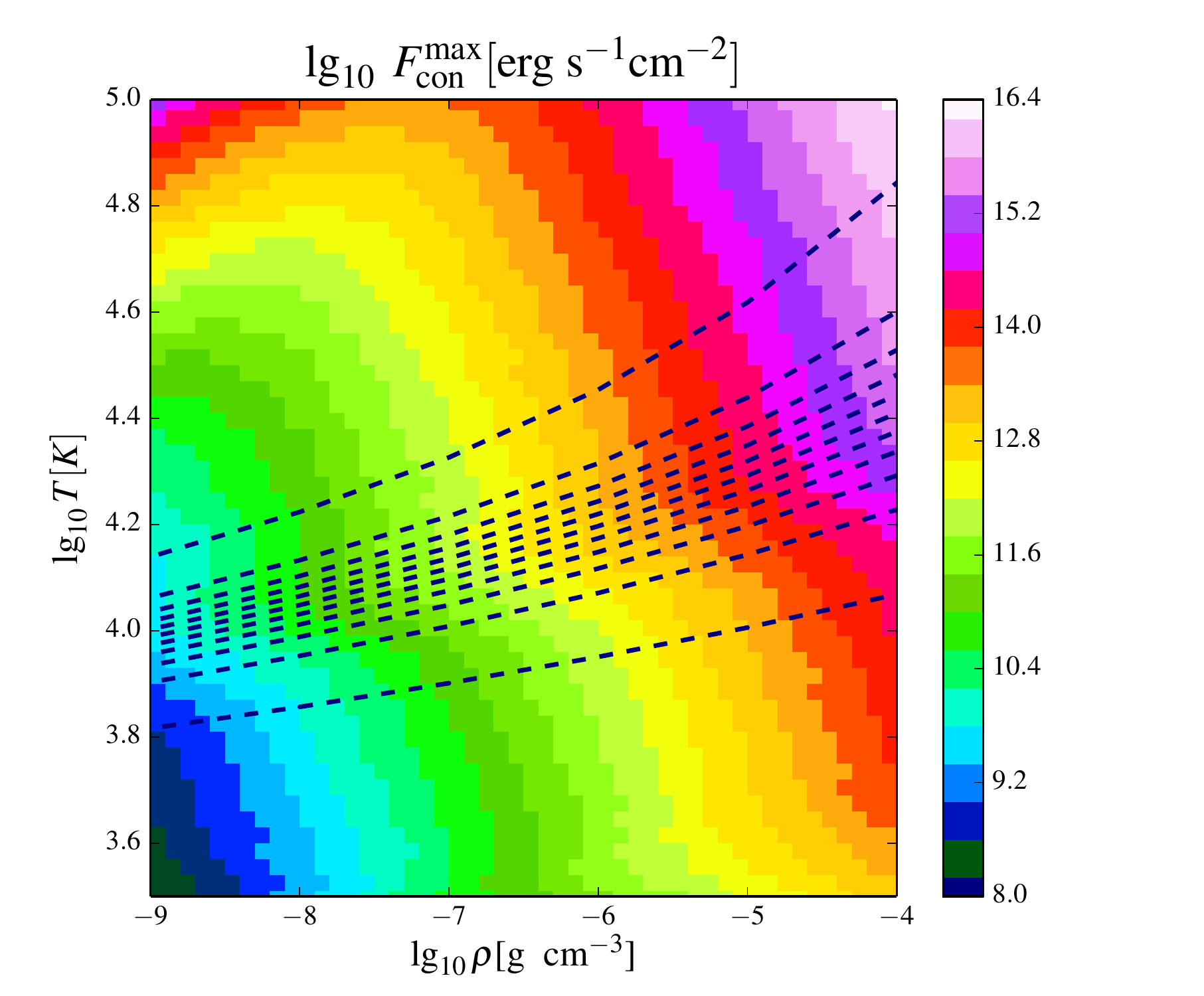}
        \caption{The maximum possible convective flux for a stellar mixture with X=0.7 and Y=0.28 (see Equation \ref{eq:fcon_max}). Lines for Hydrogen partial ionization zones are as in Figure \ref{fig:op_ion}.}
    \label{fig:con_flux}
\end{figure}
\begin{figure}
        \includegraphics[width=3.6in]{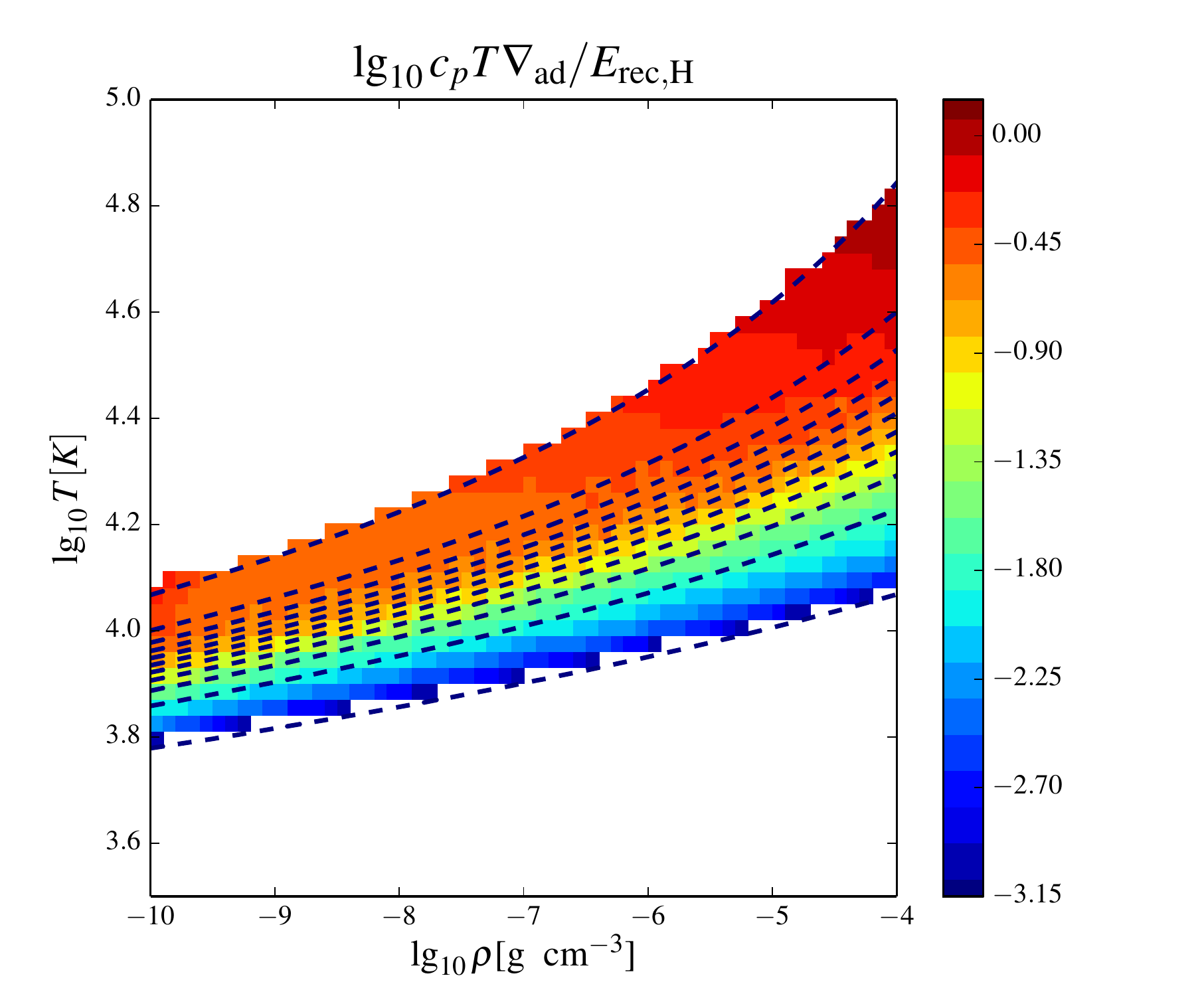}
        \caption{This plotted quantity characterizes the ratio between the maximum possible convective flux and the flux created by local recombination, assuming that at each location $F_{\rm con}^{\rm max}$ has to carry the recombination energy released by changing the ionization fraction by 0.1, as between the shown contours.
        }
    \label{fig:ratio_flux}
\end{figure}

Let us search if there are partially ionized regions where radiation might dominate over convection.
First, in Figure~\ref{fig:op_ion} we show opacities in the $\rho-T$ plane for a typical stellar mixture with a mass fraction of hydrogen $X=0.7$ and mass fraction of helium $Y=0.28$. We also show there the contours of $H$ and $He$ partial ionisation, as well as typical profiles of several giant stars. Typical values of opacities within the zone of partially ionized $H$ for stellar models are $\kappa > 10$ cm$^2$ g$^{-1}$. It is those high opacity values that are expected to be the main cause of the weakness  of the radiative flux to move energy, as compared to the convective flux.

In the regions where convection is very efficient, $({\nabla-\nabla_e})$ could become as small  as $\sim 0.01$. However, in the regions where convection competes with radiation, and the convection itself is not very effective, the ratio $\nabla/(\nabla-\nabla_e) \rightarrow 1$. As convective flux struggles to transfer energy away, it responds by boosting the local convective velocity (this is a consequence of the buoyancy force increasing when $(\nabla-\nabla_e)$ is not small any more), and $\Mach_{cv} \rightarrow 1$. We introduce $f_{\rm ratio}(\rho,T)$ such that 

\begin{equation}
  \label{eq:ratio_flux}
  \frac{F_{\rm rad}}{F_{\rm conv}} \approx f_{\rm ratio} (\rho,T)   \frac{g}{10^{-4} g_\odot}  
\end{equation}
\begin{equation}
  f_{\rm ratio} (\rho,T) =  \frac{4 a c  }{3} \frac{T^3}{ \kappa P \rho c_s c_p}   \ 10^{-4} g_\odot 
  \label{eq:fratio}
\end{equation}

\noindent In Figure~\ref{fig:fratio} we show $f_{\rm ratio} (\rho,T)$ for a typical stellar mixture (X=0.7, Y=0.28).  Please note that while we analyze the case of all giant stars, the location of the border between where each of the fluxes dominates is to the left of the most expanded stars. It therefore makes sense to scale to the case $g = 10^{-4} g_\odot$. It can be seen that the radiative flux is far less efficient than the convective flux everywhere where hydrogen ionisation takes place at densities above  $\rho_{\rm crit}\sim 2\times 10^{-10}$ g cm$^{-3}$. Even changing $\Mach_{cv}$ to 0.1 would affect results for $\rho_{\rm crit}$ by just a bit.
It is discussed that in the case of luminous massive stars, the effective local opacity in a three-dimensional star can be lower than that predicted by a one-dimensional model, due to the effect of the porosity \citep{2004ApJ...616..525O,2015ApJ...813...74J}. This effect may potentially play a role for $f_{\rm ratio}$.  We note that the effect of porosity needed to make radiation more effective than convection, for the region of interest, requires changing opacities by two orders of magnitude or more.

Radiation is capable of removing recombination energy only from a region where it is more efficient than convection. \cite{2017MNRAS.472.4361S} have argued that a consideration of a random photon walk and the comparison of time that it takes for a photon to escape can lead to most of the released hydrogen recombination energy to be transported away. While the photons that are carried by radiation can indeed be taken away, their contribution to the overall energy transport from the zones of partial recombination is proportional to the ratio of the fluxes. Hence, the optical depth of the recombination zone, as well as the photon diffusion time and its comparison to a CE timescale, do not matter, so long as the convective flux dominates.

Only in cases when the envelope's material, while expanding and cooling, passes through the recombination region while having density less than $\rho_{\rm crit}$, can radiation take the recombination energy away. To constrain which stars can do that, let us consider the envelope's material evolution during a CEE. The track of a fluid element on the $\rho-T$ plane can be split into three stages. First, there is an entropy increase due to either frictional orbital energy deposition or shocks. Second is the stage of adiabatic expansion while the plasma is still fully ionized. And finally, an expansion stage while recombining; this is accompanied by a decrease of plasma entropy. The track on the $\rho-T$ plan goes to smaller densities and smaller temperatures from its original location for all three stages.

We introduce the scaled entropy as $s=S/(k_B N_A)$, where $k_B$ is the Boltzmann constant and $N_A$ is the Avogadro number. We show values of scaled entropy in Figure~\ref{fig:entr}. One can see that the profiles of the stars in convective regions are aligned along the lines of constant entropy while wiggling in zones of partial ionisation, as expected.

To transport energy by radiation from the recombination region, i.e., to start the recombination at densities comparable to $\rho_{\rm crit}$, the scaled entropy of the material prior to the adiabatic expansion has to be more than 40 mol g$^{-1}$. Only very expanded donors could have an entropy of their convective envelopes that large prior to a CEE.

During a CEE, stellar material is expected to get an entropy boost, at the onset of its adiabatic expansion. What does that entropy boost mean? For an ionized gas, a boost by one unit in scaled entropy is equivalent to providing to the same material a specific heat $k_B N_A T$. At the same time, its specific internal energy at the same temperature is $3/(2\mu) k_B N_A T$. Note that an initially bound envelope implies that its internal energy is roughly half of the absolute value of its specific potential energy. Providing more than  $\sim 2.5 k_B N_A T$ makes the envelope material immediately unbound. Therefore, if a stellar model prior to the onset of the CEE has its scaled entropy below about 37 mol  g$^{-1}$, radiation is not expected to remove the recombination energy.

\section{Ability of the convective flux to remove the recombination energy}

Let us rewite Equation~\ref{eq:fconv} as 

\begin{equation}
  F_{\rm con} =  \Mach_{\rm cv} \frac{(\nabla - \nabla_e)}{\nabla_{\rm ad}} F_{\rm con}^{\rm max} , \
\end{equation}

\noindent where $F_{\rm con}^{\rm max} $ is the ``maximum possible convective flux''

\begin{equation}
  F_{\rm con}^{\rm max} =  \rho c_s c_p T {\nabla_{\rm ad}} \ .
  \label{eq:fcon_max}
\end{equation}

\noindent Note that this maximum possible convective flux assumes that the convective velocity is the same as the local sonic velocity.
In principle, an MLT-equipped stellar code could not produce a valid result for convective energy transport in this regime,
but we still can use it as the upper limit. Conveniently, the quantity $F_{\rm con}^{\rm max} $  depends only on the EOS (see Figure \ref{fig:con_flux}).
Numerically, we find that we can approximate that, for most of the region of  interest, the limiting flux $F_{\rm con}^{\rm max}\propto \rho^{0.8} T^{1.3}$.

Let us consider a fully ionized fluid element that is initially located above the hydrogen recombination region on the $\rho-T$ plane. Assume that it receives some energy (an entropy boost) from the shrinking binary orbit, and starts its expansion. Figure \ref{fig:con_flux} shows that along an expansion track  the value of $F_{\rm con}^{\rm max}$ for that fluid element decreases. During an adiabatic expansion of an ideal ionized gas, $T \propto \rho ^{2/3}$, resulting in $F_{\rm con}^{\rm max}\propto \rho^{5/3}$. For a self-similar expansion, $\rho \propto r^{-3}$. Then $L\propto  \Mach_{\rm cv} {(\nabla - \nabla_e)} F_{\rm con}^{\rm max} r^2 \propto  \Mach_{\rm cv} {(\nabla - \nabla_e)} r^{-3}$ -- as the CE expands, the local luminosity has to drop, unless  both $\Mach_{\rm cv}$ and $(\nabla - \nabla_e)$ strongly increase, but their values are limited. The initial expansion of a CE, to some extent, becomes self-driving. This self-driving regime will break down once the thermal timescale of the expanding envelope becomes comparable to the dynamical timescale, and the adiabatic regime of the expansion stops, or when recombination starts to play a key role and an adiabatic expansion with $\gamma=5/3$ is no longer a valid approximation.

Let us now consider the recombination and convection.  Hydrogen recombination provides specific energy $E_{\rm rec, H}= 1.3\times 10^{13} X f_h$~erg~g$^{-1}$,
where $X$ is the hydrogen mass fraction and $f_h=H^+/H$ is how much of the hydrogen is ionized. Now let us assume that recombination takes place on the same timescale as the fluid element moves through the ionisation region, the recombination timescale $\tau_{\rm rec}$. We can relate it to the local dynamical timescale  $\tau_{\rm dyn}$ as $\alpha_{\rm rec}=\tau_{\rm rec}/\tau_{\rm dyn}$. For dynamical CEEs, $\alpha_{\rm rec}\approx 1$, while for self-regulated CEEs,  $\alpha_{\rm rec}\ga 10$. The width of the hydrogen recombination zone can be written as $dr_{\rm rec}\approx \alpha_{\rm H} r$.
For giants, the width of the recombination zone is a substantial part of their total radius, with $\alpha_{\rm H}\approx 0.1-0.5$, and $\alpha_{\rm H}$ only increases as the CE expands. The local energy flux that is created due to recombination then can be found as 

\begin{equation}
F_{\rm rec} = \frac{E_{\rm rec, H}}{\alpha_{\rm rec} \tau_{\rm dyn}} \rho dr = \frac{\alpha_{\rm H}}{\alpha_{\rm rec}} E_{\rm rec, H} \frac{ v_{\rm esc}}{\sqrt{2}}
\end{equation}

\noindent Here $v_{\rm esc}=\sqrt{2Gm/r}$ is the local escape speed.
We can form the ratio between the maximum possible local convective flux 
and the energy flux due to recombination:

\begin{equation}
\frac{F_{\rm con}^{\rm max} }{F_{\rm rec} } = \frac{\sqrt{2}c_s}{v_{\rm esc}} \frac{c_p T \nabla_{\rm ad}}{E_{\rm rec, H}} \frac{\alpha_{\rm rec}}{\alpha_{\rm H}}
\end{equation}  

\noindent The maximum possible convective flux can remove the released recombination energy only if that ratio is more than one; otherwise, the shell will expand. The first term, $\sqrt{2}c_s/v_{\rm esc}$ is about one for any star that is stable at the on-set of a CE. The quantity $c_p T \nabla_{\rm ad}/E_{\rm rec, H}$ depends only on the EOS and is plotted in Figure~\ref{fig:ratio_flux}, only for the part of the $\rho-T$ plane where the recombination energy can be released. As can be seen, the convective flux is trapped in the low ionisation regions. For example, if $f_{\rm H}<0.2$, to take away the energy from the recombination zone, either the timescale for the envelope expansion has to be up to 100 times larger than the local dynamical timescale, or the width of the recombining zone $\alpha_{\rm H}$ has to be less than 0.01. Please note that only on short, dynamical timescales the recombination energy cannot be removed by fluxes. This is why stars evolving from the main  sequence to the   red giant phase  do not lose their envelopes when  they go through hydrogen   and helium recombinations.

Note that in giants, prior to the CE, convection is strongly subsonic. An increase of the convective velocity from subsonic to sonic speeds implies that part of the released recombination energy goes into the kinetic energy of convective eddies. For typical subsonic convection prior to the CE, the energy that becomes kinetic energy is about $c_s^2/2= \Gamma_1/ (2 \mu) k_B N_A T$. I.e., if a fluid element has to be sped up to sonic velocity, its specific kinetic energy becomes comparable to its specific internal energy (note that a slightly supersonic fluid element is becoming locally unbound). It will store, therefore, most of the released recombination energy as kinetic energy. Stellar codes usually ignore the kinetic energy contained in the convective eddies, as they do not carry much energy in the case of subsonic convection, nor is the acceleration of the convective eddies a process which converts thermal energy into kinetic energy on a dynamical timescale. But for CEEs this effect has to be taken into account, at least when the ability of the covective flux to carry energy is considered.

\section{Conclusions}

We have considered whether radiative or convective flux is capable of removing the energy that is released during hydrogen recombination during regular CEEs. We limit ourselves to regular (dynamic) CEEs, since self-regulated CEEs, similar to, for example, those considered in \cite{2018arXiv180305864G}, cannot in principle provide insight into the efficiency of recombination energy usage, due to their intrinsic feature of freezing the envelope recombination profile in order to match the constant energy source.

For regular CEEs, we have shown that the convective flux is incapable of removing the released recombination energy even in the limit of transonic  convection. As a result, the released recombination energy is fated to be utilized as a work term. Also, for convection to become transonic or faster, it  has to convert most of the released recombination energy into kinetic energy. In both cases, through the work term, or via kinetic energy boost, the recombination energy is used locally helping to expand and eject the CE. We anticipate as well that if stellar codes that are used to model CEE show that the convection has to be transonic or faster, the results of the calculations might be invalid, as in this case the MLT is used in a regime beyond its validity.

On the other hand, the radiative flux can potentially only transfer the recombination energy from donors where, prior to CE, their scaled entropy is $s\ga 37$ mol g$^{-1}$. In donors with a lower value of entropy in their envelopes, envelope material does not pass through the region where radiative transfer is effective enough to remove the recombination energy; for them convection always dominates. For those donors, if their envelope receives an entropy boost  large enough to allow the matter to start recombination in the radiation dominated regime, the envelope material will become unbound before the recombination starts. An evaluation of the recombination energy losses by radiative diffusion, whether by photon diffusion time, or by the optical depth of the recombination zone, should not be done when the radiative transfer itself is not dominating the overall energy transfer.

\acknowledgments

We thank the unknown referee for useful comments. 
NI  acknowledges support  from  CRC program,   funding from  NSERC
Discovery, and acknowledges  that a  part  of  this work  was
performed at  the KITP  which is  supported in part  by the  NSF under
Grant  No.  NSF PHY17-48958.

%\bibliographystyle{aasjournal}
%\bibliography{recomb} 

\end{document}